\documentclass[final,3p,times,onecolumn]{elsarticle}
\usepackage{amssymb}
\usepackage[fleqn]{amsmath}
\usepackage{amsthm}
\usepackage{subfigure}
\usepackage{graphicx}

\numberwithin{equation}{section}
\biboptions{numbers,sort&compress}
\usepackage{hyperref}
\usepackage{times}
\usepackage{anysize}
\marginsize{2.5cm}{2.5cm}{1cm}{2cm}
\begin{document}

\begin{frontmatter}

\title{Non-local symmetries of the Hirota-Satsuma coupled KdV system and their applications}

\author{ Junchao Chen \fnref{label2} }
\author{ Xiangpeng Xin \fnref{label2} }
\author{ Yong Chen \fnref{label2} \corref{cor1} }
\ead{ychen@sei.ecnu.edu.cn}

\cortext[cor1]{Corresponding author. Shanghai Key Laboratory of Trustworthy Computing, East China Normal University, Shanghai, 200062, People's Republic of China}

\address[label2]{ Shanghai Key Laboratory of Trustworthy Computing, East China Normal University, Shanghai, 200062, People's Republic of China}

\begin{abstract}
%% Text of abstract
The nonlocal symmetry is derived from the known Darboux transformation (DT) of the Hirota-Satsuma
coupled KdV (HS-cKdV) system, and infinitely many nonlocal symmetries are obtained by introducing some internal parameters.
By extending the HS-cKdV system to an auxiliary system with five dependent variables, the prolongation is found to localize the nonlocal symmetry related to the DT.
Base on the enlarged system, the finite symmetry transformations and similarity reductions about the local symmetries are computed, which lead to some novel exact solutions of the HS-cKdV system.
These solutions contain some new solutions from old ones by the finite symmetry
transformation and exact interaction solutions among solitons and other complicated waves including
periodic cnoidal waves and Painlev\'{e} waves through similarity reductions. Some integrable models from the nonlocal symmetry related to the DT are obtained in two aspects:
the negative HS-cKdV hierarchy obtained by introducing the internal parameter and integrable models both in lower and higher dimensions given by restricting the symmetry constraints.

\end{abstract}

\end{frontmatter}

%%
%% Start line numbering here if you want
%%
% \linenumbers

%% main text
\section{Introduction}

Symmetry study is always a powerful method in physics and other natural and applied
sciences, especially, in integrable systems and solion theory \cite{m-olver-1993,m-bluman-1989}.
For one known symmetry of a differential system, there are some important applications, such as
obtaining new solutions from old ones via corresponding finite transformation \cite{m-olver-1993,m-bluman-1989,j-gf-1992-25}, reducing dimensions of differential system by
similarity reductions \cite{m-olver-1993,m-bluman-1989,j-ck-1989-30,j-ychen-na-2009,j-ychen-zna-2009} and getting new integrable hierarchies and higher dimensional integrable models from original integrable models \cite{j-lou-1997-27,j-lou-plb-1993-261,j-lou-jpa-1997-301,j-lou-ps-1998-57} and so on.
A lot of work has devoted to finding the general Lie symmetry by using the classical or
nonclassical Lie group method in the past years.
However, since it is still challengeable to seek
nonlocal symmetries for a given differential system \cite{j-gga-jpa-1993-26,j-gga-jmp-1993-26,j-gga-prsla-1994-446}, only a few papers have paid attention
to the relevant research.
Most recently, some efficient techniques for searching nonlocal symmetries of differential equations have been proposed and developed. For example, one can obtained infinite many nonlocal symmetries by recursion
operators and their inverses \cite{j-gga-prsla-1994-446,j-lou-ijmpa-1993-3,j-lou-pla-1993-175,j-lou-jmp-1994-2336,j-lou-jmp-1994-2390},  the conformal invariant form (Schwartz form) \cite{j-lou-jpa-1997-301,j-lou-ps-1998-57},  Darboux transformation (DT) \cite{j-lou-jpa-1997-30,j-lou-jmp-1997-38,j-hlc-pre-2012-85,j-xing-2012-arxiv}, B\"{a}cklund transformation \cite{j-lou-jpa-2012-45}, pseudopotentials \cite{j-gf-1992-25} and potential system \cite{m-bluman-1989} and so on.

However, the finite symmetry transformations and similarity reductions cannot be directly calculated from the nonlocal symmetry. This fact leads to few works to construct explicit
solutions for the related differential equation(s) in this aspect. Fortunately, the localization of nonlocal symmetry through introducing potential \cite{m-bluman-1989} and pseudopotential-type \cite{ m-edg-1980,j-kv-aam-1984-2,j-kv-aam-1989-15}
symmetries which possess close prolongation extends the applicability of symmetry method to obtain some new solutions of differential equations. Base on this method, new finite symmetry transformations can be derived (for example, the finite symmetry transformations obtained from DT and the initial DT are different but possess same infinitesimal form \cite{j-hlc-pre-2012-85}) and the original
equation(s) need to be expanded into some prolonged systems \cite{j-gf-1992-25,j-hlc-pre-2012-85,j-lou-jpa-2012-45}. Further, to solve these prolonged systems generates some novel exact interaction solutions among solitons and other complicated waves including
periodic cnoidal waves and Painlev\'{e} waves \cite{j-hlc-pre-2012-85,j-lou-jpa-2012-45}. Practically, these novel solutions unearthed have potentially important applications in some physical fields \cite{j-shj-jpa-2004-37,j-shj-jpa-2005-38,j-shj-pre-2005-71}.

Moreover, to find new integrable models is another important application of
symmetry study.
For (1+1)-dimensional integrable
model, the recursion operator \cite{j-ow-pla-1983,j-fb-ptp-1982-68,j-gks-jmp-1999-40} is a valid method to produce the integrable hierarchy.
Very recently, Lou \cite{j-lou-ijmpa-1993-3,j-lou-pla-1993-175,j-lou-jmp-1994-2336,j-lou-jmp-1994-2390} has extended some negative hierarchy through utilizing the inverse recursion operator for (1+1)-dimensional integrable
models and introducing some internal parameters for (2+1)-dimensional integrable
models \cite{j-lou-ps-1998-57}. In addition, making use of the nonlinearization of Lax pair under certain constraints
between potentials and eigenfunctions, Cao has developed a systematic approach to find finite-dimensional
integrable systems \cite{j-cao-sca-1990-33,j-cao-ams-1991-7,j-cao-jmp-1999-40}. In particular, it has also been pointed
that by restricting a symmetry constraint to the Lax pair of the soliton equation, one can not
only obtain the lower dimensional integrable models from higher ones, but can also embed the
lower ones into higher dimensional integrable models \cite{j-lou-jpa-1997-301,j-lou-jmp-1997-38,j-lou-ctp-1996-25}.

In this paper, we concentrate on the nonlocal symmetry of the Hirota-Satsuma coupled KdV (HS-cKdV) system and their applications.
Firstly,  the nonlocal symmetry is derived from the known DT of the HS-cKdV system, and infinitely many nonlocal symmetries are obtained by introducing some internal parameters.
The prolongation of the nonlocal symmetry related to the DT is found by extending the HS-cKdV system to an auxiliary system with five dependent variables. Then, the finite symmetry
transformations and similarity reductions about the corresponding local symmetries are computed for giving novel exact solutions of the HS-cKdV system.
These solutions contain some new solutions from old ones by the finite symmetry
transformations and exact interaction solutions among solitons and other complicated waves including
periodic cnoidal waves and Painlev\'{e} waves by similarity reductions. The another work of the paper
is to extend the HS-cKdV system to some new integrable models from the nonlocal symmetry related to the DT in two aspects:
the negative HS-cKdV hierarchy obtained by introducing the internal parameter and some other integrable models both in finite
and infinite dimensions given by restricting the symmetry constraints.

The HS-cKdV system
\begin{eqnarray}
\label{hs-01} &&u_t=\frac{1}{2}u_{xxx}+3uu_x-6vv_x,\\
\label{hs-02} &&v_t=-v_{xxx}-3uv_x,
\end{eqnarray}
is  proposed as the coupled KdV system by Hirota and Satsuma, which describes interactions of two long waves
with different dispersion relations \cite{j-hr-pla-1981}.
In a following paper \cite{j-hr-jpsj-1982}, these authors shown that this coupled KdV system is the four-reduction of the
celebrated KP hierarchy and its soliton solutions can be derived from ones of the KP equation.
Meanwhile, Wilson \cite{j-wg-pla-1982} observed that the HS-cKdV system is just an example of many integrable systems arose from the
Drinfeld-Sokolov theory \cite{j-dvg-jsm-1985,b-gvg-1981}.
Some significant properties of the HS-cKdV system have been revealed in the past years. For instance,
the HS-cKdV system possesses bilinear form \cite{j-hr-pla-1981,j-twh-pla-2000}, Lax pair \cite{j-dr-pla-1982,j-wj-jmp-1984,j-wj-jmp-1985}, B${\rm \ddot{a}}$cklund transformations \cite{j-ld-pla-1983}, Darboux transformations \cite{j-lsb-jpa-1993,j-hhc-csf-2003,j-hhc-pla-2008}, Painlev${\rm \acute{e}}$ property \cite{j-wj-jmp-1984,j-wj-jmp-1985}, infinitely many
symmetries and conservation laws \cite{j-ow-pla-1983} etc.

The paper is organized as follows. In Sec.II, the nonlocal
symmetry is derived from the DT of the HS-cKdV system, and more nonlocal symmetries can be produced from one seed symmetry through introducing some inner parameters.
A prolonged system to localize the nonlocal symmetry is presented by extending the HS-cKdV system. The finite symmetry
transformations of prolonged local symmetries and similar reductions of the prolonged system are presented, and some new exact solutions of the original system are obtained.
Section III is devoted to finding the negative HS-cKdV hierarchy by introducing the internal parameter and integrable models both in lower and higher dimensions by restricting the symmetry constraints. The last
section are conclusions and discussions.

\section{Non-local symmetries via Darboux transformation}

It is known that DT is the most direct and yet elementary approach for the construction of exact solutions.
Using this method, one can obtain new solutions fron old solutions through simple iteration.
In this section, we use the invariant properties of differential equations exhibited by DT to deduce the nonlocal symmetries of the HS-cKdV system (\ref{hs-01})-(\ref{hs-02}).

The Lax pair for the HS-cKdV system (\ref{hs-01})-(\ref{hs-02}) reads \cite{j-dr-pla-1982,j-wj-jmp-1984,j-wj-jmp-1985}
\begin{eqnarray}
\label{hs-03}  \psi_{1xx} &=& -(u+v)\psi_1-\lambda\psi_2, \\
\label{hs-04}  \psi_{2xx} &=& -(u-v)\psi_2+\lambda\psi_1, \\
\label{hs-05}  \psi_{1t} &=& -\frac{1}{2}(u_x-2v_x)\psi_1+(u-2v)\psi_{1x}-2\lambda\psi_{2x}, \\
\label{hs-06}  \psi_{2t} &=& -\frac{1}{2}(u_x+2v_x)\psi_2+(u+2v)\psi_{2x}+2\lambda\psi_{1x},
\end{eqnarray}
where $\{u,v \}$ is a solution of Eqs.(\ref{hs-01})-(\ref{hs-02}) and $\lambda$ is a spectral parameter.

In Ref.\cite{j-hhc-csf-2003,j-hhc-pla-2008}, Hu and Liu had constructed the DT for the HS-cKdV system from singularity analysis and reduction of a binary DT. We rewrite it here.

\emph{Proposition 1}\cite{j-hhc-csf-2003,j-hhc-pla-2008} \ \ \ The DT of Eqs.(\ref{hs-01})-(\ref{hs-02}) is expressed by
\begin{equation}\label{hs-07}
  \bar{u}=u+2(\ln\theta)_{xx},\ \ \bar{v}=v+\frac{\psi_2\psi_{1x}-\psi_{1}\psi_{2x}}{\theta},
\end{equation}
with
\begin{equation}\label{hs-08}
    \theta_x=\psi_1\psi_2,\ \ \theta_t=2\lambda(\psi^2_1-\psi^2_2)-2\psi_{1x}\psi_{2x}-u\psi_1\psi_2.
\end{equation}

\emph{Proposition 2}
\begin{eqnarray}
\label{hs-09}  \sigma_1 =(\sigma^u_1,\sigma^v_1) \equiv \bigg(\Big(\frac{\theta_2}{\theta_1}\Big)_{xx} , \frac{W[\phi_2,\tilde{\phi}_1]-W[\phi_1,\tilde{\phi}_2]}{\theta_1}+\frac{\theta_2W[\phi_1,\phi_2]}{\theta^2_1}\bigg)
\end{eqnarray}
 is a symmetry of the HS-cKdV system (\ref{hs-01})-(\ref{hs-02}) with $\{u,v\}$ replaced by $\{U,V\}$, where $\phi_1,\phi_2,\tilde{\phi}_1,\tilde{\phi}_2$ and $\theta_1,\theta_2$
satisfy the following equations:
%\begin{eqnarray}
%% \nonumber to remove numbering (before each equation)
%  &&\phi_{1xx} = -(U-2(\ln\theta_1)_{xx}+V-\frac{\phi_2\phi_{1x}-\phi_1\phi_{2x}}{\theta_1})\phi_1, \\
%  &&\phi_{2xx} = -(U-2(\ln\theta_1)_{xx}-V+\frac{\phi_2\phi_{1x}-\phi_1\phi_{2x}}{\theta_1})\phi_2, \\
%  &&\phi_{1t} = -\frac{1}{2}\bigg((U-2(\ln\theta_1)_{xx})_x-2(V-\frac{\phi_2\phi_{1x}-\phi_1\phi_{2x}}{\theta_1})_x\bigg)\phi_1 +((U-2(\ln\theta_1)_{xx})-2(V-\frac{\phi_2\phi_{1x}-\phi_1\phi_{2x}}{\theta_1}))\phi_{1x},\\
%  &&\phi_{2t} = -\frac{1}{2}\bigg((U-2(\ln\theta_1)_{xx})_x+2(V-\frac{\phi_2\phi_{1x}-\phi_1\phi_{2x}}{\theta_1})_x\bigg)\phi_2 +((U-2(\ln\theta_1)_{xx})+2(V-\frac{\phi_2\phi_{1x}-\phi_1\phi_{2x}}{\theta_1}))\phi_{2x},
%\end{eqnarray}
\begin{eqnarray}
% \nonumber to remove numbering (before each equation)
\label{hs-10}  &&\phi_{1xx} = -(\bar{U}+\bar{V})\phi_1, \ \ \phi_{1t} = -\frac{1}{2}(\bar{U}_x-2\bar{V}_x)\phi_1 +(\bar{U}-2\bar{V})\phi_{1x}, \\
\label{hs-11}  &&\phi_{2xx} = -(\bar{U}-\bar{V})\phi_2, \ \ \phi_{2t} = -\frac{1}{2}(\bar{U}_x+2\bar{V}_x)\phi_2 +(\bar{U}+2\bar{V})\phi_{2x},\\
\label{hs-12}  &&\tilde{\phi}_{1xx} = -(\bar{U}+\bar{V})\tilde{\phi}_1-\phi_2, \ \ \tilde{\phi}_{1t} = -\frac{1}{2}(\bar{U}_x-2\bar{V}_x)\tilde{\phi}_1 +(\bar{U}-2\bar{V})\tilde{\phi}_{1x}-2\phi_{2x},\\
\label{hs-13} &&\tilde{\phi}_{2xx} = -(\bar{U}-\bar{V})\tilde{\phi}_2+\phi_1, \ \ \tilde{\phi}_{2t} = -\frac{1}{2}(\bar{U}_x+2\bar{V}_x)\tilde{\phi}_2 +(\bar{U}+2\bar{V})\tilde{\phi}_{2x}+2\phi_{1x},
\end{eqnarray}
and
\begin{eqnarray}
% \nonumber to remove numbering (before each equation)
\hspace{-2cm} \label{hs-14} && \theta_{1}=\int\phi_1\phi_2dx+\alpha(t),\ \ \theta_{1t}=-2\phi_{1x}\phi_{2x}-\bar{U}\phi_1\phi_2, \\
\hspace{-2cm} \label{hs-15} && \theta_{2}=\int\phi_1\tilde{\phi}_2+\phi_2\tilde{\phi}_1dx+\beta(t),\ \ \theta_{2t}=2(\phi^2_1-\phi^2_2)-2(\phi_{1x}\tilde{\phi}_{2x}+\phi_{2x}\tilde{\phi}_{1x})-\bar{U}(\phi_1\tilde{\phi}_2+\phi_2\tilde{\phi}_1),
\end{eqnarray}
where $\bar{U}\equiv U-2(\ln\theta_1)_{xx}, \bar {V}\equiv V-\frac{[\phi_2,\phi_1]}{\theta_1}$, $W[a,b] \equiv ab_x-ba_x$, and $\{\alpha(t),\beta(t)\}$ are functions of $t$.

\emph{Proof}. Setting $\phi_1 {\equiv} \psi_1(x,t,0)$, $\phi_2 {\equiv} \psi_2(x,t,0)$, $\tilde{\phi}_1 {\equiv} \psi_{1\lambda}(x,t,0)$, $\tilde{\phi}_{2} {\equiv} \psi_{2\lambda}(x,t,0)$, $\theta_1 {\equiv} \theta(x,t,0)$ and $\theta_2 {\equiv} \theta_{\lambda}(x,t,0)$.
Furthermore, defining $U=u+2(\ln\theta_1)_{xx}$ and $V=v+\frac{[\phi_2,\phi_1]}{\theta_1}$. From proposition 1, we know that $U,V$ is a
solution of the HS-cKdV system (\ref{hs-01}) and (\ref{hs-02}). Now we formally expand $\bar{u}$ and $\bar{v}$ in powers of $\lambda$. We obtain
\begin{eqnarray}
\label{hs-16}  \bar{u} &=& U+ \lambda \bigg[\Big(2\frac{\partial^2}{\partial x^2}\ln\psi\Big)_{\lambda} \bigg | _{\lambda =0 } \bigg]+\textrm{O}(\lambda^2)=U+2\lambda\Big(\frac{\theta_2}{\theta_1}\Big)_{xx}+\textrm{O}(\lambda^2), \\
\nonumber  \bar{v}&=& V+ \lambda \bigg[\Big(\frac{W[\psi_2,\psi_{1}]}{\theta}\Big)_{\lambda}\bigg | _{\lambda =0 }\bigg]+\textrm{O}(\lambda^2)\\
\label{hs-17} &=&V+\lambda \bigg(\frac{W[\phi_2,\tilde{\phi}_1]-W[\phi_1,\tilde{\phi}_2]}{\theta_1}+\frac{\theta_2W[\phi_1,\phi_2]}{\theta^2_1}\bigg)+\textrm{O}(\lambda^2).
\end{eqnarray}
Thus $(\sigma_1,\sigma_2)$ is a symmetry of HS-cKdV system (\ref{hs-01}) and (\ref{hs-02}) with respect to $U$ and $V$. Then, substituting $u=U-2(\ln\theta_1)_{xx}$ and $v=V-\frac{[\phi_2,\phi_1]}{\theta_1}$ in
(\ref{hs-01}) and (\ref{hs-02}), one can derive (\ref{hs-10})-(\ref{hs-15}). Finally, we have completed the proof of \emph{proposition 2}.

Furthermore, a direct calculation shows that if $\phi_1$ and $\phi_2$  satisfies (\ref{hs-10}) and (\ref{hs-11}), then
\begin{eqnarray}
% \nonumber to remove numbering (before each equation)
\label{hs-18}  \tilde{\phi}_1 &=& -\phi_1 \int\Big[\frac{1}{\phi^2_1}\int \phi_1\phi_2dx\Big]dx+F_1\phi_1 \int\frac{1}{\phi^2_1}dx +F_2\phi_1, \\
\label{hs-19}  \tilde{\phi}_2 &=& -\phi_2 \int\Big[\frac{1}{\phi^2_2}\int \phi_2\phi_2dx\Big]dx+G_1\phi_2 \int\frac{1}{\phi^2_2}dx +G_2\phi_2,
\end{eqnarray}
is a solution of (\ref{hs-12}) and (\ref{hs-13}), where $F_1,F_2,G_1$ and $G_2$ are arbitrary constants.

On the other hand, from Eqs.(\ref{hs-10}) and (\ref{hs-11}), one can find such a fact that if $\{\phi_1,\phi_2\}$ is a solution of (\ref{hs-10}) and (\ref{hs-11}), then
$\{\phi_1,\phi_2\}$ has the following relation to Lax pairs (\ref{hs-03})-(\ref{hs-05}) with $\lambda=0$:
\begin{eqnarray}
\label{hs-20} \phi_1=-A_1\int\frac{1}{\psi^2_1}+A_2,\ \ \phi_2=A_3\frac{\psi_2}{\psi_1}\int\frac{1}{\psi^2_2}+A_4\frac{\psi_2}{\psi_1},
\end{eqnarray}
with the constraint,
\begin{equation}
 \label{hs-21}  A_3\int\frac{1}{\psi^2_1}\int\frac{1}{\psi^2_2}+A_1A_4\int\frac{1}{\psi^2_1}-A_2A_3\int\frac{1}{\psi^2_2}-A_2A_4=\frac{\psi_{1x}}{\psi_1\psi_2},
\end{equation}
where $A_1,A_2,A_4$ and $A_4$ are arbitrary constants.

Now, substituting (\ref{hs-14}),(\ref{hs-15}) and (\ref{hs-18})-(\ref{hs-21}) in the symmetry (\ref{hs-09}), we obtain a different nonlocal symmetry of the HS-cKdV system corresponding
to Lax pairs (\ref{hs-03})-(\ref{hs-05}). In other words, the nonlocal symmetry we obtained finally is exhibited by $\psi_1$, $\psi_2$ in Eqs.(\ref{hs-03})-(\ref{hs-05}).
The final expression of this nonlocal symmetry is very complicated, we omit it here. However, we find that many arbitrary constants and functions with respect to $t$ exist in the last form of the symmetry derived. Therefore, we only concern about part of our result, which still is a nonlocal symmetry
of the original equations. So we select the coefficient of $\beta(t)$ in the complex expression of this symmetry for the further investigation in the following paper.
We have

\emph{Proposition 3}\ \ \
$\sigma_2=(\sigma^u_2,\sigma^v_2)$ is a nonlocal symmetry of HS-cKdV system (\ref{hs-01})-(\ref{hs-02}), where
\begin{eqnarray}
 \label{hs-22} \nonumber  \sigma^u_2 &=& 2A_1A_3\Lambda_1\int\frac{1}{\psi^2_1}\int\frac{1}{\psi^2_2}+2A_1(A_4\Lambda_1+A_3\frac{\psi_1}{\psi_2})\int\frac{1}{\psi^2_1}-2A_3(A_2\Lambda_1-A_1\frac{\psi_2}{\psi_1})\int\frac{1}{\psi^2_2}\\
 &&-2A_2A_4\Lambda_1+2A_1A_4\frac{\psi_2}{\psi_1}-2A_2A_3\frac{\psi_1}{\psi_2},\\
 \label{hs-23} \nonumber\sigma^v_2 &=& -A_1A_3\Lambda_2\int\frac{1}{\psi^2_1}\int\frac{1}{\psi^2_2}-A_1(A_4\Lambda_2+A_3\frac{\psi_1}{\psi_2})\int\frac{1}{\psi^2_1}+A_3(A_2\Lambda_2+A_1\frac{\psi_2}{\psi_1})\int\frac{1}{\psi^2_2}\\
 &&+A_2A_4\Lambda_2+A_1A_4\frac{\psi_2}{\psi_1}+A_2A_3\frac{\psi_1}{\psi_2},
\end{eqnarray}
and $\Lambda_1=(\psi_1\psi_2)_x$, $\Lambda_2=W[\psi_1,\psi_2]$, meanwhile $\psi_1,\psi_2$ satisfy Eqs.(\ref{hs-03})-(\ref{hs-06}) with $\lambda=0$.

Moreover, it is easily verified the fact that Lax pairs (\ref{hs-03})-(\ref{hs-06}) is invariant under the transformations:
\begin{equation}\label{hs-24}
   \psi_1 \rightarrow \bar{\psi}_1=\psi_1\int\frac{1}{\psi^2_1}dx, \ \ \psi_2 \rightarrow \bar{\psi}_2=\psi_2\int\frac{1}{\psi^2_2}dx.
\end{equation}
With the aid of the above transformations (\ref{hs-24}), the nonlocal symmetry $\sigma_2$ in Proposition 3 can be rewritten by the form:
$-A_1A_3\Lambda(\bar{\psi}_1,\bar{\psi}_2)-A_1A_4\Lambda(\bar{\psi}_1,\psi_2)+A_2A_3\Lambda(\psi_1,\bar{\psi}_2)+A_2A_4\Lambda(\psi_1,\psi_2)$
with $\Lambda=(2\Lambda_1,\Lambda_2)$. Considering the inverse transformations of (\ref{hs-24}), one can easily know

\emph{Proposition 4} \ \ \ If $\psi_1,\psi_2$ satisfy Lax pair in Eqs.(\ref{hs-03})-(\ref{hs-06}) with $\lambda=0$, then
\begin{equation}\label{hs-25}
   \sigma_3=(\sigma^u_3,\sigma^v_3) \equiv \Big(-2(\psi_1\psi_2)_x, \psi_1\psi_{2x}-\psi_2\psi_{1x}\Big)
\end{equation}
is a seed symmetry of the HS-cKdV system (\ref{hs-01})-(\ref{hs-02}).

\emph{Remark 1}\ \  In fact, if $\psi_1,\psi_2$ satisfy Lax pair in Eqs.(\ref{hs-03})-(\ref{hs-06}) with the arbitrary spectral parameter $\lambda$,
$\sigma_3$ in (\ref{hs-25}) is still a symmetry of the HS-cKdV system. This fact is easily verified by direct calculation.

In Ref.\cite{j-lou-ctp-1996-25,j-lou-jmp-1997-38}, the author presented one kind of method to seek for more symmetries via differentiating a known one with respect to inner parameters:

\emph{Proposition 5} \ \ \ If a $\lambda$-dependent function $\sigma_0(\lambda)$ is a symmetry of the HS-cKdV system (\ref{hs-01})-(\ref{hs-02}) with $\lambda{\equiv}\{\lambda_1,\lambda_2,...\lambda_r\}$,
then
\begin{equation}
  \sigma_{n} \equiv \frac{d^{\{n\}}}{d\lambda^{\{n\}}}\sigma_0(\lambda)\equiv \frac{d^{\{n_1\}}}{d\lambda_1^{\{n_1\}}}\frac{d^{\{n_2\}}}{d\lambda_2^{\{n_2\}}} \cdot\cdot\cdot\frac{d^{\{n_r\}}}{d\lambda_r^{\{n_r\}}}\sigma_0(\lambda)
\end{equation}
is also a symmetry of the same HS-cKdV system (\ref{hs-01})-(\ref{hs-02}) for $\{n\}{\equiv}\{n_1, n_2, \cdots, n_r\}$.

Using the Proposition 4 and Proposition 5, we can get a set of infinitely many new nonlocal symmetries. For instance, if we suppose $\{\psi_1,\psi_2\}$ and $\{\hat{\psi}_1,\hat{\psi}_2\}$ are two solutions of the Lax pair in Eqs.(\ref{hs-03})-(\ref{hs-05}), then
\begin{eqnarray*}
 \sigma(\lambda_1,\lambda_2)\equiv\Big(\sigma^u(\lambda_1,\lambda_2),\sigma^v(\lambda_1,\lambda_2)\Big)
\end{eqnarray*}
with
\begin{eqnarray*}
&&\sigma^u(\lambda_1,\lambda_2)=-2[(\lambda_1\psi_1+\lambda_2\hat{\psi}_1)(\lambda_1\psi_2+\lambda_2\hat{\psi}_2)]_x, \\ &&\sigma^v(\lambda_1,\lambda_2)=(\lambda_1\psi_1+\lambda_2\hat{\psi}_1)(\lambda_1\psi_2+\lambda_2\hat{\psi}_2)_x-(\lambda_1\psi_2+\lambda_2\hat{\psi}_2)(\lambda_1\psi_1+\lambda_2\hat{\psi}_1)_x,
\end{eqnarray*}
and $\frac{\partial^{n_1+n_2}}{\partial\lambda_1^{n_1}\partial\lambda_2^{n_2}}\sigma(\lambda_1,\lambda_2)$ are also symmetries of the HS-cKdV system (\ref{hs-01})-(\ref{hs-02}).

\section{Localization of the non-local symmetry and exact solutions}

\subsection{Localization of the non-local symmetry from DT}
In this section, we devote to study the symmetry from DT given by Eq.(\ref{hs-25}) in above section. First, we rewrite it as
\begin{equation}\label{hs-26}
  \sigma^{u}=-2\psi_{1}\psi_{2x}-2\psi_2\psi_{1x},\ \ \sigma^{v}=\psi_1\psi_{2x}-\psi_2\psi_{1x},
\end{equation}
which apparently contain derivative terms $\psi_{1x}$ and $\psi_{2x}$. Hence, by
introducing new dependent variables $\varphi_1 {\equiv} \varphi_1(x,t)$ and $\varphi_2 {\equiv} \varphi_2(x,t)$ with
\begin{equation}\label{hs-27}
   \varphi_1=\psi_{1x},\  \  \ \varphi_2=\psi_{2x},
\end{equation}
the above symmetry (\ref{hs-26}) is converted into
\begin{equation}\label{hs-28}
   \sigma^{u}=-2\psi_{1}\varphi_2-2\psi_2\varphi_1,\ \ \sigma^{v}=\psi_1\varphi_2-\psi_2\varphi_1.
\end{equation}

In the following, we shall list the linearized equations of Eqs. (\ref{hs-03})-(\ref{hs-04}) and (\ref{hs-27}) with $\lambda=0$,
\begin{eqnarray}
\nonumber &&  \sigma^{\psi_1}_{xx}+(u+v)\sigma^{\psi_1}+(\sigma^u+\sigma^v)\psi_1=0, \\
\label{hs-29} &&  \sigma^{\psi_2}_{xx}+(u-v)\sigma^{\psi_2}+(\sigma^u-\sigma^v)\psi_2=0, \\
\nonumber &&  \sigma^{\psi_1}_x-\sigma^{\varphi_1}=0,\ \ \sigma^{\psi_2}_x-\sigma^{\varphi_2}=0.
\end{eqnarray}
where $\sigma^u$ and $\sigma^v$ is given by (\ref{hs-28}) and $\sigma^{\psi_1},\sigma^{\psi_2},\sigma^{\varphi_1}$ and $\sigma^{\varphi_2}$ denote
the symmetries of $\psi_1, \psi_2, \varphi_1$ and $\varphi_2$, respectively.

Observing from Eq.(\ref{hs-03}) and (\ref{hs-04}), one can easily deduce that solution of (\ref{hs-29}) has the form:
\begin{eqnarray}
\label{hs-30}&&  \sigma^{\psi_1}=p\psi_1,\ \ \sigma^{\psi_2}=p\psi_2, \\
\label{hs-31}&&  \sigma^{\varphi_1}=\psi^2_1\psi_2+p\varphi_1,\ \ \sigma^{\varphi_2}=\psi_1\psi^2_2+p\varphi_2.
\end{eqnarray}
where $p{\equiv}p(x,t)$ is a new potential variable which we introduced to make the prolonged system close completely,
and it satisfies identically the compatibility conditions:
\begin{equation}
\label{hs-32}    p_x=\psi_1\psi_2,\ \ p_t=-u\psi_1\psi_2-2\varphi_1\varphi_2.
\end{equation}
 Furthermore, the linearized equation of its symmetry $\sigma^p$ read
 \begin{eqnarray}
\label{hs-33}  \sigma^p_x=\psi_1\sigma^{\psi_2}+\psi_2\sigma^{\psi_1}.
\end{eqnarray}
and a straightforward calculation shows that $\sigma^p$ has the simple form
\begin{eqnarray}
\label{hs-34}  \sigma_p=p^2.
\end{eqnarray}

Finally, the prolongation for nonlocal symmetry (\ref{hs-26}) is successfully localized by introducing variables $\{\psi_1,\\ \psi_2,\varphi_1,\varphi_2,p \}$ on
the original variables $\{u,v\}$ with the equivalent vector expression
\begin{eqnarray}
\label{hs-35} \nonumber V&=&-2(\psi_1\varphi_2+\psi_2\varphi_1)\frac{\partial}{\partial u}+(\psi_1\varphi_2-\psi_2\varphi_1)\frac{\partial}{\partial v}+p\psi_1\frac{\partial}{\partial \psi_1}+p\psi_2\frac{\partial}{\partial \psi_2}\\
&&+(\psi^2_1\psi_2+p\varphi_1)\frac{\partial}{\partial \varphi_1}+(\psi_1\psi^2_2+p\varphi_2)\frac{\partial}{\partial \varphi_2}+p^2\frac{\partial}{\partial p}.
\end{eqnarray}

Here, it is worthy to mention that if further consider the differential equation of the introduced variable $p$ from above localized procedure,
one can find the corresponding differential equation is nothing but the Schwartz form of the HS-cKdV system (\ref{hs-01})-(\ref{hs-02})
 \begin{eqnarray}
\label{hs-36}   3\mathcal{H}^3_x-6\mathcal{H}\mathcal{H}_x\mathcal{H}_{xx} + 4\mathcal{H}^2\Big(2\mathcal{C}_t-(\mathcal{C}^2)_x-4\mathcal{C}\mathcal{S}_x+\mathcal{H}_{xxx}-\mathcal{C}_{xxx}\Big)+16\mathcal{H}^3\mathcal{S}_x=0,
\end{eqnarray}
 where $\mathcal{H}=\mathcal{S}+2\mathcal{C}$, $\mathcal{S}=\frac{p_{xxx}}{p_x}-\frac{3p^2_{xx}}{p^2_x}$ and $\mathcal{C}=-\frac{p_t}{p_x}$ are all invariant under M\"{o}bious (conformal) transformation.
The reason lies that the finite M\"{o}bious transformation
\begin{equation*}
\nonumber p\rightarrow\frac{a+bp}{c+dp},\ \ (ad \neq bc)
\end{equation*}
possess its infinitesimal transformation $p\rightarrow p+\epsilon p^2$ in special case $a=0,b=c=1$ and $d=-\epsilon$.
The fact show us that the corresponding Schwartz form of a given differential equation derived usually by utilizing singularities analysis method
can also be obtained through localization of the nonlocal symmetry from DT. Therefore, this method may provide a potential application to find Schwartz form of
some integrable models, especially discrete integrable models which have some known DTs.

\subsection{Finite symmetry transformation}

For the related prolonged system from symmetry (\ref{hs-26}), it is natural to seek the finite transformation form of (\ref{hs-35}).
By using Lie's first theorem, we need to solve the following initial value problem
%\begin{eqnarray}
%&&\frac{d\hat{u}(\epsilon)}{d\epsilon}=-2\hat{\psi}_1\hat{\phi}_2-2\hat{\psi}_2\hat{\phi}_1, \ \ \hat{u}(0)=u,\\
%&&\frac{d\hat{v}(\epsilon)}{d\epsilon}=\hat{\psi}_1\hat{\phi}_2-\hat{\psi}_2\hat{\phi}_1, \ \  \ \hat{v}(0)=v,\\
%&&\frac{d\hat{\psi_1}(\epsilon)}{d\epsilon}=\hat{p}\hat{\psi}_1, \ \ \ \ \hat{\psi}_1(0)=\psi_1, \\
%&&\frac{d\hat{\psi_2}(\epsilon)}{d\epsilon}=\hat{p}\hat{\psi}_2, \ \ \ \ \hat{\psi}_2(0)=\psi_2, \\
%&&\frac{d\hat{\phi_1}(\epsilon)}{d\epsilon}=\hat{\psi}^2_1\hat{\psi}_2+\hat{p}\hat{\phi}_1, \ \ \ \ \hat{\phi}_1(0)=\phi_1, \\
%&&\frac{d\hat{\phi_2}(\epsilon)}{d\epsilon}=\hat{\psi}_1\hat{\psi}^2_2+\hat{p}\hat{\phi}_2, \ \ \ \ \hat{\phi}_2(0)=\phi_2, \\
%&&\frac{d\hat{p}(\epsilon)}{d\epsilon}=\hat{p}^2,\ \ \ \ \ \hat{p}(0)=p,
%\end{eqnarray}
\begin{eqnarray}\label{hs-37}
 \nonumber &&\frac{d\hat{u}(\epsilon)}{d\epsilon}=-2\hat{\psi}_1\hat{\varphi}_2-2\hat{\psi}_2\hat{\varphi}_1, \ \ \frac{d\hat{v}(\epsilon)}{d\epsilon}=\hat{\psi}_1\hat{\varphi}_2-\hat{\psi}_2\hat{\varphi}_1, \ \ \frac{d\hat{\psi_1}(\epsilon)}{d\epsilon}=\hat{p}\hat{\psi}_1, \frac{d\hat{\psi_2}(\epsilon)}{d\epsilon}=\hat{p}\hat{\psi}_2, \\
 &&  \frac{d\hat{\varphi_1}(\epsilon)}{d\epsilon}=\hat{\psi}^2_1\hat{\psi}_2+\hat{p}\hat{\varphi}_1,\ \
\frac{d\hat{\phi_2}(\epsilon)}{d\epsilon}=\hat{\psi}_1\hat{\psi}^2_2+\hat{p}\hat{\varphi}_2,\ \ \frac{d\hat{p}(\epsilon)}{d\epsilon}=\hat{p}^2, \\
 \nonumber &&    \hat{u}(0)=u,\ \ \hat{v}(0)=v,\ \ \hat{\psi}_1(0)=\psi_1,\ \ \hat{\psi}_2(0)=\psi_2,\ \ \hat{\varphi}_1(0)=\varphi_1,\ \ \hat{\varphi}_2(0)=\varphi_2,\ \ \hat{p}(0)=p,
\end{eqnarray}
where $\epsilon$ is the group parameter.

By solving the initial value problem (\ref{hs-37}), we arrive at the symmetry group theorem as follow:

\textbf{Theorem 1} If $\{u,v,\psi_1,\psi_2,\varphi_1,\varphi_2,p\}$ is the solution of the extended system consist of (\ref{hs-01})-(\ref{hs-06}), (\ref{hs-27}) and $(\ref{hs-32})$ with $\lambda=0$, so is
$\{\hat{u},\hat{v},\hat{\psi}_1,\hat{\psi}_2,\hat{\varphi}_1,\hat{\varphi}_2,\hat{p}\}$
\begin{eqnarray}\label{hs-38}
\nonumber && \hat{u}=u-\frac{2\epsilon(\psi_1\varphi_2+\psi_2\varphi_1)}{1-\epsilon p}-\frac{2\epsilon^2\psi^2_1\psi^2_2}{(1-\epsilon p)^2}, \\
\nonumber && \hat{v}=v+\frac{\epsilon(\psi_1\varphi_2+\psi_2\varphi_1)}{1-\epsilon p}+\frac{\epsilon^2\psi^2_1\psi^2_2}{(1-\epsilon p)^2},\\
&& \hat{\psi}_1=\frac{\psi_1}{1-\epsilon p},\ \ \hat{\psi}_2=\frac{\psi_2}{1-\epsilon p},\\
\nonumber && \hat{\varphi}_1=\frac{\varphi_1}{1-\epsilon p}+\frac{\epsilon\psi_2\psi^2_1}{(1-\epsilon p)^2},\ \ \hat{\varphi}_2=\frac{\varphi_2}{1-\epsilon p}+\frac{\epsilon\psi_1\psi^2_2}{(1-\epsilon p)^2},\\
\nonumber && \hat{p}=\frac{p}{1-\epsilon p},
\end{eqnarray}
with $\epsilon$ is arbitrary group parameter.

For example, starting from the simple solution $\{u=\frac{\lambda^2_1}{2}-\frac{\lambda^2_2}{2},v=\frac{\lambda^2_2}{2}-\frac{\lambda^2_1}{2} \}$ of (\ref{hs-01}) and (\ref{hs-02}), we can derive the corresponding special solutions for the introduced dependent variables from (\ref{hs-03})-(\ref{hs-06}), (\ref{hs-27}) and $(\ref{hs-32})$ under the condition $\lambda=0$,
\begin{eqnarray}\label{hs-39}
\nonumber&&\psi_1=\cosh\Big(\lambda_1x+\frac{\lambda_1(\lambda^2_1-3\lambda^2_2)}{2}t+\xi_{10}\Big),\\
\nonumber&&\psi_2=\cosh\Big(\lambda_2x+\frac{\lambda_2(\lambda^2_2-3\lambda^2_1)}{2}t+\xi_{20}\Big),\\
&&\phi_1=\lambda_1\sinh\Big(\lambda_1x+\frac{\lambda_1(\lambda^2_1-3\lambda^2_2)}{2}t+\xi_{10}\Big),\\
\nonumber&&\phi_2=\lambda_2\sinh\Big(\lambda_2x+\frac{\lambda_2(\lambda^2_2-3\lambda^2_1)}{2}t+\xi_{20}\Big),\\
\nonumber&& p=\frac{\sinh\xi_1}{2\lambda_1-2\lambda_2}+\frac{\sinh\xi_2}{2\lambda_1+2\lambda_2}+\xi_0.
\end{eqnarray}
Substituting (\ref{hs-39}) into (\ref{hs-38}) leads to the non-trivial solution of HS-cKdV system (\ref{hs-01})-(\ref{hs-02})
{\small
\begin{eqnarray}\label{hs-40}
\hspace{-2cm} \nonumber&&u=-\frac{\lambda^2_1}{2}-\frac{\lambda^2_2}{2}+\frac{\epsilon[(\lambda_1-\lambda_2)\sinh\xi_1+(\lambda_1+\lambda_2)\sinh\xi_2]}{\epsilon(\frac{\sinh\xi_1}{2\lambda_1-2\lambda_2}+\frac{\sinh\xi_2}{2\lambda_1+2\lambda_2}+\xi_0)-1}
-\frac{\epsilon^2(\cosh\xi_1+\cosh\xi_2)^2}{16[\epsilon(\frac{\sinh\xi_1}{2\lambda_1-2\lambda_2}+\frac{\sinh\xi_2}{2\lambda_1+2\lambda_2}+\xi_0)-1]^2},\\
\hspace{-2cm} &&v=-\frac{\lambda^2_1}{2}+\frac{\lambda^2_2}{2}-\frac{\epsilon[(\lambda_1-\lambda_2)\sinh\xi_1+(\lambda_1+\lambda_2)\sinh\xi_2]}{2\epsilon(\frac{\sinh\xi_1}{2\lambda_1-2\lambda_2}+\frac{\sinh\xi_2}{2\lambda_1+2\lambda_2}+\xi_0)-2}
+\frac{\epsilon^2(\cosh\xi_1+\cosh\xi_2)^2}{32[\epsilon(\frac{\sinh\xi_1}{2\lambda_1-2\lambda_2}+\frac{\sinh\xi_2}{2\lambda_1+2\lambda_2}+\xi_0)-1]^2},
\end{eqnarray}}
with
\begin{eqnarray*}
&&\xi_1=(\lambda_1-\lambda_2)[x+\frac{\lambda^2_1+4\lambda_1\lambda_2+\lambda^2_2}{2}t]+\xi_{10}-\xi_{20},\\
&& \xi_2=(\lambda_1+\lambda_2)[x+\frac{\lambda^2_1-4\lambda_1\lambda_2+\lambda^2_2}{2}t]+\xi_{10}+\xi_{20},
\end{eqnarray*}
where $\xi_0,\xi_{10}$ and $\xi_{20}$ are arbitrary constants.

It is necessary to point out that the finite transformation exhibited in \textbf{Theorem 1} is distinct from the original DT in above section.
Indeed, it is equivalent to the so-called Levi transformation, or the second type of Darboux transformations. At the algebra level, two different types of
finite transformation possesses the same infinitesimal expression (\ref{hs-26}).
Besides, the last equation of (\ref{hs-38}) is nothing but the corresponding M\"{o}bious transformation in the front analysis about Schwartz form (\ref{hs-36}) of the original system.

\subsection{Similarity reductions of the prolonged system}

%For the closed prolonged system, it is worthy to further study some significant properties.
In this section, our main aim is to seek for some exact solutions from the prolonged system to construct some novel solutions of original HS-cKdV system.
Therefore, we employ the classical Lie symmetry method to search for similarity reductions of the whole prolonged system.

Accordingly, we consider the one-parameter
group of infinitesimal transformations in $\{x,t,u$,$v$,$\psi_1$,\\ $\psi_2$,$\varphi_1$,$\varphi_2$,$p\}$ given by
\begin{eqnarray*}
 &&\{x,t,u,v,\psi_1,\psi_2,\varphi_1,\varphi_2,p\} {\rightarrow} \{x{+}\epsilon X,t{+}\epsilon T,u{+}\epsilon U,v{+}\epsilon V,\psi_1{+}\epsilon \psi_1,\psi_2{+}\epsilon\Psi_2,\varphi_1{+}\epsilon\Phi_1,\varphi_2{+}\epsilon \Phi_2,p{+}\epsilon P\}
\end{eqnarray*}
with
\begin{eqnarray}\label{hs-41}
\nonumber && \sigma^{u}=X u_x + T u_t -U,\ \ \sigma^{v}=X v_x + T v_t -V,\\
\nonumber&& \sigma^{\psi_1}=X \psi_{1x} + T \psi_{1t} -\Psi_1,\ \ \sigma^{\psi_2}=X \psi_{2x} + T \psi_{2t} -\Psi_2,\\
&& \sigma^{\varphi_1}=X \varphi_{1x} + T \varphi_{1t} -\Phi_1,\ \ \sigma^{\varphi_2}=X \varphi_{2x} + T \varphi_{2t} -\Phi_2,\\
\nonumber&& \sigma^{p}=X p_x + T p_t -P,
\end{eqnarray}
where $X,T,U,V,\Psi_1,\Psi_2,\Phi_1,\Phi_2$ and $P$ are functions with respect to $\{x,t,u,v,\psi_1,\psi_2,\varphi_1,\varphi_2,p\}$, and $\epsilon$ is a small parameter.
Substituting (\ref{hs-41}) into the symmetry equations, i.e., the linearized equations of the prolonged system obtained in above section
\begin{eqnarray}\label{hs-42}
\hspace{-2cm} \nonumber  && \sigma^{u}_t-\frac{1}{2}\sigma^{u}_{xxx}-3u\sigma^{u}_x-3\sigma^{u}u_x+6v\sigma^{v}_x+6\sigma^{v}v_x=0,\ \ \sigma^{v}_t+\sigma^{v}_{xxx}+3\sigma^{u}v_x+3u\sigma^{v}_x=0\\
 \nonumber && \sigma^{\psi_1}_{xx}+(u+v)\sigma^{\psi_1}+(\sigma^u+\sigma^v)\psi_1=0,\ \ \sigma^{\psi_2}_{xx}+(u-v)\sigma^{\psi_2}+(\sigma^u-\sigma^v)\psi_2=0, \\
 \nonumber && \sigma^{\psi_1}_t+\frac{1}{2}\sigma^{\psi_1}(u_x-2v_x)+\frac{1}{2}\psi_1(\sigma^{u}_x-2\sigma^{v}_x)-(u-2v)\sigma^{\psi_1}_x-(\sigma^{u}-2\sigma^{v})\psi_{1x}=0,\\
 && \sigma^{\psi_2}_t+\frac{1}{2}\sigma^{\psi_2}(u_x+2v_x)+\frac{1}{2}\psi_2(\sigma^{u}_x+2\sigma^{v}_x)-(u+2v)\sigma^{\psi_2}_x-(\sigma^{u}+2\sigma^{v})\psi_{2x}=0,\\
 \nonumber && \sigma^{\psi_1}_x-\sigma^{\varphi_1}=0,\ \ \sigma^{\psi_2}_x-\sigma^{\varphi_2}=0,\\
\nonumber && \sigma^{p}_x-\psi_1\sigma^{\psi_2}-\psi_2\sigma^{\psi_1}=0,\ \ \sigma^{p}_t+u(\psi_1\sigma^{\psi_2}+\psi_2\sigma^{\psi_1})+\sigma^{u}\psi_1\psi_2+2\phi_1\sigma^{\varphi_2}+2\phi_2\sigma^{\varphi_1}=0,
\end{eqnarray}
then collecting together the coe¡Àcients of the dependent variables and their partial
derivatives, and setting all of them to zero, yields a system of overdetermined, linear
equations for the infinitesimals $\{X,T,U,V,\Psi_1,\Psi_2,\Phi_1,\Phi_2,P\}$.
By solving these equations, one can get
\begin{eqnarray}\label{hs-43}
\nonumber &&X=c_1x+c_4,\ \ T=3c_1t+c_2,\\
\nonumber &&U=-2c_1u-2c_3(\psi_1\varphi_2+\psi_2\varphi_1),\ \ V=-2c_1v+c_3(\psi_1\phi_2-\psi_2\phi_1),\\
&&\Psi_1=c_3p\psi_1+c_5\psi_1,\ \ \Psi_2=c_3p\psi_2+c_6\psi_2,\\
\nonumber &&\Phi_1=c_3(\psi^2_1\psi_2+p\varphi_1)-(c_1{-}c_5)\varphi_1,\ \ \Phi_2=c_3(\psi_1\psi^2_2+p\varphi_2)-(c_1{-}c_6)\varphi_2,\\
\nonumber &&P=c_3p^2+(c_5+c_6)p+c_1p+c_7,
\end{eqnarray}
where $c_i$ $(i{=}1...7)$ are arbitrary constants. Especially, when $c_1=c_2=c_4=c_5=c_6=c_7=0$, the degenerated symmetry is just one (\ref{hs-35}),
and when $c_3=c_5=c_6=c_7=0$, the related symmetry is only the general Lie point symmetry of the HS-cKdV system (\ref{hs-01})-(\ref{hs-02}).

To give some corresponding group invariant solutions, we need to solve the following characteristic equations:
\begin{equation}\label{hs-44}
\frac{dx}{X}=\frac{dt}{T}=\frac{du}{U}=\frac{dv}{V}=\frac{d\psi_1}{\Psi_1}=\frac{d\psi_2}{\Psi_2}=\frac{d\varphi_1}{\Phi_1}=\frac{d\varphi_2}{\Phi_2}=\frac{dp}{P}.
\end{equation}
Next, we consider several different similarity reductions arising from (\ref{hs-44}) under the condition $c_3\neq 0$ in some detail.

\textbf{Reduction 1} $c_1\neq 0$

Without loss of generality, we assume $c_2=c_4=c_5=c_6=0$ and redefine $k^2=\frac{c^2_1-4c_3c_7}{36c^2_1}$. Then, two different situations, $k \neq 0$ and $k=0$, need to be further considered, respectively.

\emph{Case 1} $k\neq0$. We obtain
similarity solutions
\begin{eqnarray}\label{hs-45}
\nonumber  && u=\frac{U(z)}{t^{\frac{2}{3}}}+\frac{2c_3}{9c^2_1k^2t^{\frac{2}{3}}}\exp(-\frac{2}{3}P(z)) \{3c_1k[\Psi_1(z)\Phi_2(z){+}\Psi_2(z)\Phi_1(z)]\tanh z_1{-}c_3\Psi^2_1(z)\Psi^2_2(z){\rm sech}^2z_1\},\\
\nonumber  && v=\frac{V(z)}{t^{\frac{2}{3}}}+\frac{c_3 }{3c_1kt^{\frac{2}{3}}}\exp(-\frac{2}{3}P(z))[\Psi_1(z)\Phi_2(z)-\Psi_2(z)\Phi_1(z)]\tanh z_1,\\
 &&\psi_1=\frac{\Psi_1(z)}{t^{\frac{1}{6}}}\exp(-\frac{1}{6}P(z)){\rm sech}z_1,\ \ \psi_2=\frac{\Psi_2(z)}{t^{\frac{1}{6}}}\exp(-\frac{1}{6}P(z)){\rm sech}z_1,\\
\nonumber  &&\varphi_1=\frac{1}{3c_1k t^{\frac{1}{2}}}\exp(-\frac{1}{2}P(z))[3c_1k\Phi_1(x){\rm sech}z_1 + c_3\Psi^2_1(z)\Psi_2(z)\tanh z_1],\\
\nonumber  &&\varphi_2=\frac{1}{3c_1kt^{\frac{1}{2}}}\exp(-\frac{1}{2}P(z))[3c_1k\Phi_2(x){\rm sech}z_1 + c_3\Psi_1(z)\Psi^2_2(z)\tanh z_1],\\
\nonumber  && p=-\frac{c_1}{2c_3}(1+6k\tanh z_1),
\end{eqnarray}
with $z_1=k(\ln t+P(z))$, and the similarity variable $z=x/\sqrt[3]{t}$.

Substituting (\ref{hs-45}) into the prolonged equations leads to
\begin{eqnarray}\label{hs-46}
\nonumber &&\hspace{-2.3cm} U(z)=-\frac{\Psi_{1zz}(z)}{2\Psi_1(z)}-\frac{\Psi_{2zz}(z)}{2\Psi_2(z)}-\frac{c_3}{9c_1k^2}\exp(-\frac{1}{3}P(z))[\Psi_1(z)\Psi_{2z}(z)+\Psi_2(z)\Psi_{1z}(z)]\\
\nonumber &&\hspace{-1.2cm}        -\frac{c^2_3(1+12k^2)}{108c^2_1k^4}\exp(-\frac{2}{3}P(z))\Psi^2_1(z)\Psi^2_2(z),\\
 &&\hspace{-2.3cm} V(z)=-\frac{\Psi_{1zz}(z)}{2\Psi_1(z)}+\frac{\Psi_{2zz}(z)}{2\Psi_2(z)}+\frac{c_3}{18c_1k^2}\exp(-\frac{1}{3}P(z))[\Psi_1(z)\Psi_{2z}(z)-\Psi_2(z)\Psi_{1z}(z)],\\
\nonumber &&\hspace{-2.3cm} \Phi_1=\exp(\frac{1}{3}P(z))\Psi_1(z)+\frac{c_3}{18c_1k^2}\Psi^2_1(z)\Psi_2(z),\ \ \Phi_2=\exp(\frac{1}{3}P(z))\Psi_2(z)+\frac{c_3}{18c_1k^2}\Psi_1(x)\Psi^2_2(z),\\
\nonumber &&\hspace{-2.3cm} \Psi_1(z)=\sqrt{-\frac{3c_1k^2}{c_3}P_z(z)}\exp(\frac{1}{6}P(z)+Q(z)),\ \ \Psi_2(z)=\sqrt{-\frac{3c_1k^2}{c_3}P_z(z)}\exp(\frac{1}{6}P(z)-Q(z)),
\end{eqnarray}
and $P_z(z)\equiv P_1(z)$, $Q_z(z)\equiv Q_1(z)$ satisfy ordinary differential equations
\begin{eqnarray}\label{hs-47}
 \nonumber &&\hspace{-1.5cm} 6P_{1zz}(z)P_1(z)-9P^2_{1z}(z)-12P_1(z)+4P^2_1(z)z+36P^2_1(z)Q^2_1(z)-12k^2P^4_1(z)=0,\\
 &&\hspace{-1.5cm}  3Q_{1zz}P_1(z)-4P_1(z)Q_1(z)z+9Q_1(z)-24P_1(z)Q^3_1(z)=0.
\end{eqnarray}

%By using the ARS algorithm, we know that equations (\ref{hs-47}) allow the singularities $P_1(z)=P_{10}/(z-z_0)$ and $Q_1(z)=Q_{10}/(z-z_0)$ with $P_{10}=\pm\frac{1}{k}$ and $Q_{10}=\pm\frac{1}{2}$, and the resonant points occurs at $\{-2,-1,2,4\}$. Then detailed calculation show equations (\ref{hs-47}) can pass the Painleve test.

By using the ARS algorithm, eliminating $P_1(z)$ and its derivative terms in first equation through the second equation in Eq.(\ref{hs-47}), we obtain a four-order ordinary differential equation about the variable
$Q_1(z)$. Then, we get two possible branches: $Q_1(z)=Q_{10}/(z-z_0)$ with $Q_{10}{=}\{\pm\frac{1}{2}\}$,  and the resonant points occurs at $\{-1,1,4,5\}$. Then detailed calculation show equations (\ref{hs-47}) can pass the Painlev\'{e} test.

\emph{Case 2} $k=0$. We obtain similarity solutions
\begin{eqnarray}\label{hs-48}
\nonumber &&\hspace{0cm} u=\frac{U(z)}{t^{\frac{2}{3}}}+\frac{2c_3}{3c_1 t^{\frac{2}{3}}z_1}[\Psi_1(z)\Phi_2(z)+\Psi_2(z)\Phi_1(z)]-\frac{2c^2_3}{9c^2_1t^{\frac{2}{3}}z^2_1}\Psi^2_1(z)\Psi^2_2(z),\\
&&\hspace{0cm} v=\frac{V(z)}{t^{\frac{2}{3}}}-\frac{c_3}{3c_1 t^{\frac{2}{3}}z_1}[\Psi_1(z)\Phi_2(z)-\Psi_2(z)\Phi_1(z)],\\
\nonumber &&\hspace{0cm} \phi_1=\frac{\Phi_1(z)}{t^{\frac{1}{2}}z_1}-\frac{c_3\Psi^2_1(z)\Psi_2(z)}{3c_1t^{\frac{1}{2}}z^2_1},\ \ \phi_2=\frac{\Phi_2(z)}{t^{\frac{1}{2}}z_1}-\frac{c_3\Psi_1(z)\Psi^2_2(z)}{3c_1t^{\frac{1}{2}}z^2_1},\\
\nonumber &&\hspace{0cm} \psi_1=\frac{\Psi_1(z)}{t^{\frac{1}{6}}z_1},\ \ \psi_2=\frac{\Psi_2(z)}{t^{\frac{1}{6}}z_1},\ \ p=-\frac{c_1}{2c_3}-\frac{3c_1}{c_3z_1},
\end{eqnarray}
with $z_1=\ln t+P(z)$, and the similarity variable $z=x/\sqrt[3]{t}$.

Substituting (\ref{hs-48}) into the prolonged equations leads to
\begin{eqnarray}\label{hs-49}
 \nonumber && U(z)=-\frac{\Psi_{1zz}(z)}{2\Psi_1(z)}-\frac{\Psi_{2zz}(z)}{2\Psi_2(z)},\ \ V(z)=-\frac{\Psi_{1zz}(z)}{2\Psi_1(z)}+\frac{\Psi_{2zz}(z)}{2\Psi_2(z)},\\
 &&\Phi_1=\Psi_{1z}(z),\ \ \Psi_1(z)=\sqrt{\frac{3c_1}{c_3}P_z(z)}\exp(Q(z)),\\
\nonumber  && \Phi_2=\Psi_{2z}(z), \ \ \Psi_2(z)=\sqrt{\frac{3c_1}{c_3}P_z(z)}\exp(-Q(z)),
\end{eqnarray}
and $P_z(z)\equiv P_1(z)$, $Q_z(z)\equiv Q_1(z)$ satisfy ordinary differential equations
\begin{eqnarray}\label{hs-50}
 \nonumber &&\hspace{0cm} 6P_{1zz}(z)P_1(z)-9P^2_{1z}(z)-12P_1(z)+4P^2_1(z)z+36P^2_1(z)Q^2_1(z)=0,\\
 &&\hspace{0cm}  3Q_{1zz}P_1(z)-4P_1(z)Q_1(z)z+9Q_1(z)-24P_1(z)Q^3_1(z)=0.
\end{eqnarray}

%For equations (\ref{hs-50}), eliminating $P_1(z)$ and its derivatives' term in first equation through the second equation, we obtain a four-order ordinary differential equation about the variable
%$Q_1(z)$. Similar to above subcase, we get four possible branches: $Q_1(z)=Q_{10}/(z-z_0)$ with $Q_{10}{=}\{\pm2,\pm\frac{1}{2}\}$,  and the resonant points appears at $\{-1,1,4,5\}$ and $\{-5,-1,4,8\}$. Then detailed calculation show equations (\ref{hs-50}) also possesses Painlev\'{e} property.
For equations (\ref{hs-50}), eliminating $P_1(z)$ and its derivative terms in first equation through the second equation, we obtain a four-order ordinary differential equation about the variable
$Q_1(z)$. Similar to above subcase, we get two possible branches: $Q_1(z)=Q_{10}/(z-z_0)$ with $Q_{10}{=}\{\pm\frac{1}{2}\}$,  and the resonant points appears at $\{-1,1,4,5\}$. Then detailed calculation show equations (\ref{hs-50}) also possesses Painlev\'{e} property.

Therefore, from the results in \textbf{Reduction 1}, one can observe that the last exact solutions of the original HS-cKdV system will include hyperbolic function
, Painlev\'{e} solution and rational function, which represent the interactions among solitary wave, Painlev\'{e} wave and rational wave.

%\begin{figure*}[!htbp]\label{hs-fig1}
%\centering
%\subfigure[]{\includegraphics[height=1.8in,width=2.3in]{fighs1-1.eps}}\hspace{0.3cm}
%\subfigure[]{\includegraphics[height=1.8in,width=2.3in]{fighs2-1.eps}}
%\subfigure[]{\includegraphics[height=1.8in,width=2.3in]{fighs1-2.eps}}\hspace{0.3cm}
%\subfigure[]{\includegraphics[height=1.8in,width=2.3in]{fighs2-2.eps}}
%\subfigure[]{\includegraphics[height=2in,width=2.5in]{fighs1-3.eps}}\hspace{0.8cm}
%\subfigure[]{\includegraphics[height=2in,width=2.5in]{fighs2-3.eps}}
%\caption{ The wave propagation plots of the HS-cKdV system given by Eq.(\ref{hs-55}), with the parameters $b_0=0.75,h=0.85$ and $m=0.65$. (a) and (b) The wave propagation pattern of the wave along $x$ axis at $t=0$; (c) and (d) The wave propagation pattern of the wave along $t$ axis at $x=0$; (e) and (f) The two-dimensional perspective view of the corresponding solution
%.}
%\end{figure*}

\textbf{Reduction 2} $c_1=0$.

For simplicity, we let $c_2=1$ and redefine the parameter $l^2{=}\frac{(c_5+c_6)^2}{4}{-}c_3c_7$. Then, two subcases $l\neq 0$ and $l=0$ are taken into account in this subsection.

\emph{Case 1} $l \neq 0$. We derive similarity solutions
\begin{eqnarray}\label{hs-51}
\nonumber && u=U(z)-\frac{2c_3}{l}[\Psi_1(z)\Phi_2(z)+\Psi_2(z)\Phi_1(z)]\tanh z_1 +\frac{2c_3}{l^2}\Psi^2_1(z)\Psi^2_2(z){\rm sech^2}z_1,\\
\nonumber && v=V(z)+\frac{c_3}{l}[\Psi_1(z)\Phi_2(z)-\Psi_2(z)\Phi_1(z)]\tanh z_1,\\
&& \psi_1=\exp(\frac{c_5-c_6}{2}t)\Psi_1(z){\rm sech}z_1,\ \ \psi_2=\exp(\frac{c_6-c_5}{2}t)\Psi_2(z){\rm sech}z_1,\\
\nonumber && \varphi_1=l^{-1}\exp(\frac{c_5-c_6}{2}t)[l\Phi_1 {\rm sech}z_1 +c_3\Psi^2_1(z)\Psi_2(z)\tanh z_1{\rm sech}z_1],\\
\nonumber && \varphi_2=l^{-1}\exp(\frac{c_6-c_5}{2}t)[l\Phi_2 {\rm sech}z_1 +c_3\Psi_1(z)\Psi^2_2(z)\tanh z_1{\rm sech}z_1],\\
\nonumber && p=-\frac{1}{2c_3}(c_5+c_6+2l\tanh z_1),
\end{eqnarray}
with $z_1=l(t+P(z))$, and the similarity variable $z=x-c_4t$.

Substituting (\ref{hs-51}) into the prolonged equations leads to
\begin{eqnarray}\label{hs-52}
 \nonumber && U(z)=-\frac{\Psi_{1zz}(z)}{2\Psi_1(z)}-\frac{\Psi_{2zz}(z)}{2\Psi_2(z)}-\frac{c^2_3}{l^2}\Psi^2_1(z)\Psi^2_2(z),\\
 \nonumber &&V(z)=-\frac{\Psi_{1zz}(z)}{2\Psi_1(z)}+\frac{\Psi_{2zz}(z)}{2\Psi_2(z)},\\
  &&\Phi_1=\Psi_{1z}(z), \ \ \Psi_1(z)=\sqrt{\frac{-l^2}{c_3P_1(z)}}\exp(\int Q_1(z)dz),\\
 \nonumber &&\Phi_2=\Psi_{2z}(z),\ \  \Psi_2(z)=\sqrt{\frac{-l^2}{c_3P_1(z)}}\exp(-\int Q_1(z)dz),\\
 \nonumber && P_1(z)= \frac{4c_4}{3}+\frac{8}{3}Q^2_1(z)-\frac{2Q_{1zz}(z)+c_5-c_6}{6Q_1(z)},
\end{eqnarray}
and $Q_1(z)$ satisfy ordinary differential equations
\begin{equation}\label{hs-53}
 Q^2_{1z}(z)= a_0+a_1Q_1(z)+a_2Q^2_1(z)+a_3Q^3_1(z)+4Q^4_1(z),
\end{equation}
with $a_0= \frac{16l^2}{a^2_3}$, $a_1=c_6-c_5$ and $a_2=4c_4$.

To show more clearly of this kind of solution, we offer one special cases of the HS-cKdV system (\ref{hs-01}) and (\ref{hs-02})  by solving Eq. (\ref{hs-53}).
For instance, A simple solution of  Eq. (\ref{hs-53}) takes the form
\begin{equation}\label{hs-54}
  Q_1(z)=b_0+b_1 {\rm sn}(hz,m),
\end{equation}
which lead to the solution of the HS-cKdV system (\ref{hs-01}) and (\ref{hs-02}):
\begin{eqnarray}\label{hs-55}
\nonumber &&\hspace{-2cm} u=-6b^2_0+\frac{1}{4}h^2(1+m^2)+\frac{(16b^4_0-h^4m^2)+2b_0mh[8b^2_0-(1+m^2)h^2]{\rm sn}(h\xi,m)}{[2b_0+mh{\rm sn}(h\xi,m)]^2} \\
 &&\hspace{-1.35cm} -\frac{\mu m h^2{\rm cn}(h\xi,m) {\rm dn}(h\xi,m) \tanh[2\mu b_0(t+z_1)]}{[2b_0+mh {\rm sn}(h\xi,m)]^2}+\frac{\mu^2 {\rm sech^2}[2\mu b_0(t+z_1)]}{2[2b_0+mh {\rm sn}(h\xi,m)]^2}, \\
\nonumber &&\hspace{-2cm} v=\frac{1}{2}\mu \tanh[2\mu b_0(t+z_1)],
\end{eqnarray}
with $z_1=\frac{1}{4b_0}\int^{\xi_0}_0[2b_0+mh {\rm sn}(h\xi,m)]^{-1}dz$, $\xi_0=x-[6b^2_0-\frac{1}{4}h^2(1+m^2)]t$ and $\mu=[(4b^2_0-h^2)(4b^2_0-m^2h^2)]^{\frac{1}{2}}$.
Here, sn, cn, and dn are usual Jacobian elliptic functions with modulus $m$.

From the expression of the last exact solution (\ref{hs-55}), we know that it potentially reflects the interaction between the soliton and the cnoidal periodic wave.
As Shin has mentioned \cite{j-shj-jpa-2004-37,j-shj-jpa-2005-38,j-shj-pre-2005-71}, these soliton$+$cnoidal wave solutions can be easily applicable to the analysis of physically interesting processes.

The dynamics behavior of the soliton$+$cnoidal wave solution given by (\ref{hs-55}) at two different choices of
the  parameters $b_0,h$ and $m$ are illustrated in Fig.1 and 2. In Fig. 1, when $m{\neq}1$, we can see that the component $u$ exhibits a bell-shaped bright soliton propagates on a cnoidal wave background,
whereas, periodic wave in the component $v$ occurs at the corner of a kink-shaped soliton. When $m{=}1$, the Jacobian elliptic periodic functions in (\ref{hs-55}) reduces to the general hyperbolic functions,
so the characteristics of the two-soltion is depicted distinctly in Fig. 2.

%\begin{figure*}[!htbp]\label{hs-fig2}
%\centering
%\subfigure[]{\includegraphics[height=2in,width=2.5in]{fighs3-3.eps}}\hspace{0.5cm}
%\subfigure[]{\includegraphics[height=2in,width=2.5in]{fighs4-3.eps}}
%\caption{ The wave propagation plots of the HS-cKdV system given by Eq.(\ref{hs-55}), with the parameters $b_0=0.8,h=0.6$ and $m=1$. (a) and (b) The two-dimensional perspective view of the corresponding solution. }
%\end{figure*}
%

\emph{Case 2}\ \ \  $l = 0$. We obtain similarity solutions
\begin{eqnarray}\label{hs-56}
\nonumber && u=U(z)+\frac{2c_3}{z_1}[\Psi_1(z)\Phi_2(z)+\Psi_2(z)\Phi_1(z)] - \frac{2c^2_3}{z^2_1}\Psi^2_1(z)\Psi^2_2(z)\\
\nonumber && v=V(z)+\frac{c_3}{z_1}[\Psi_2(z)\Phi_1(z)-\Psi_1(z)\Phi_2(z)],\\
&& \psi_1=\exp(\frac{c_5-c_6}{2}t)\frac{\Psi_1(z)}{z_1},\ \ \psi_2=\exp(\frac{c_6-c_5}{2}t)\frac{\Psi_2(z)}{z_1},\\
\nonumber && \varphi_1=\exp(\frac{c_5-c_6}{2}t)[\frac{\Phi_1(z)}{z_1}-\frac{c_3\Psi^2_1(z)\Psi_2(z)}{z^2_1}],\\
\nonumber && \varphi_2=\exp(\frac{c_6-c_5}{2}t)[\frac{\Phi_2(z)}{z_1}-\frac{c_3\Psi_1(z)\Psi^2_2(z)}{z^2_1}],\\
\nonumber && p=-\frac{c_5+c_6}{2c_3}-\frac{1}{c_3z_1},
\end{eqnarray}
with $z_1=t+P(z)$, and the similarity variable $z=x-c_4t$.

Substituting (\ref{hs-56}) into the prolonged equations leads to
\begin{eqnarray}\label{hs-57}
 \nonumber && U(z)=-\frac{\Psi_{1zz}(z)}{2\Psi_1(z)}-\frac{\Psi_{2zz}(z)}{2\Psi_2(z)},\ \ V(z)=-\frac{\Psi_{1zz}(z)}{2\Psi_1(z)}+\frac{\Psi_{2zz}(z)}{2\Psi_2(z)},\\
  &&\Phi_1=\Psi_{1z}(z), \ \ \Psi_1(z)=\sqrt{\frac{1}{c_3P_1(z)}}\exp(\int Q_1(z)dz),\\
 \nonumber &&\Phi_2=\Psi_{2z}(z),\ \  \Psi_2(z)=\sqrt{\frac{1}{c_3P_1(z)}}\exp(-\int Q_1(z)dz),\\
 \nonumber && P_1(z)= \frac{4c_4}{3}+\frac{8}{3}Q^2_1(z)-\frac{2Q_{1zz}(z)+c_5-c_6}{6Q_1(z)},
\end{eqnarray}
and $Q_1(z)$ satisfy ordinary differential equations
\begin{equation}\label{hs-58}
 Q^2_{1z}(z)= a_1Q_1(z)+a_2Q^2_1(z)+a_3Q^3_1(z)+4Q^4_1(z)
\end{equation}
with $a_1=c_6-c_5$ and $a_2=4c_4$.

Hence, in this subcase, from Eq.(\ref{hs-58}),  the final exact solutions consisted of Jacobian elliptic periodic function and rational function denote the
interactions among cnoidal periodic waves and rational waves for the HS-cKdV system.

\section{Integrable models from nonlocal symmetry}
\subsection{Negative HS-cKdV hierarchy}

As we known, the existence of infinitely many symmetry naturally leads to the existence of integrable hierarchies.
For the general hierarchies, one can obtain them form a trivial symmetry of the original integrable system using recursion operator.
For example, considering the corresponding recursion operator \cite{j-ow-pla-1983,j-fb-ptp-1982-68,j-gks-jmp-1999-40},  the HS-cKdV system has the
following high order symmetry
\begin{eqnarray}
 u_{\tau}&=&(\frac{1}{2}u_{4x}+5uu^2_x-5vv_{xx}+\frac{5}{2}u^2_{x}-8uv^2+5u^3)_x-v^2u_x,\\
 v_{\tau}&=&(-v_{4x}-2u_xv_x-6uv_{xx}-\frac{5}{3}v^3)_x-3u_{xx}v_x-5u^2v_x.
\end{eqnarray}
It is remarkable that the reduction $v=0$ in this symmetry gives us the Lax equation \cite{j-lax-cpam-1968-21}.

Recently, Lou \cite{j-lou-ijmpa-1993-3,j-lou-pla-1993-175,j-lou-jmp-1994-2336,j-lou-jmp-1994-2390} has extended some (1+1)-dimensional integrable
model to some corresponding negative hierarchy through a set of nonlocal infinitely many symmetries which
can be obtained from the kernels of a reversible recursion operator. However, to obtain the inverse of the known recursion operator is still a difficult work, especially
for the high dimensional system and the high order operator.  Here, to obtain the negative HS-cKdV hierarchy, we would like to employ one kind of method without the recursion operator \cite{j-lou-ps-1998-57}.

Starting from the nonlocal symmetry from DT given by (\ref{hs-25}), we let
\begin{equation}\label{hsnh-01}
    K^u_0(\lambda) \equiv -2(\psi_1\psi_2)_x,\ \ K^v_0(\lambda) \equiv \psi_1\psi_{2x}-\psi_2\psi_{1x},
\end{equation}
where $\psi_1$ and $\psi_2$ are determined by Lax pair (\ref{hs-03})-(\ref{hs-06}) with $\lambda\neq0$.  Because of the parameter $\lambda$ being an arbitrary
constant, we can treat it as a small parameter and expand $K^u_0(\lambda)$ and $K^v_0(\lambda)$ as a series in $\lambda$:
\begin{equation}\label{hsnh-02}
 K^u_0(\lambda)=\sum^{\infty}_{n=0}\frac{1}{n!}\frac{\partial^n}{\partial\lambda^n}K^u_0(\lambda)\bigg | _{\lambda =0 }\lambda^n,\ \
 K^v_0(\lambda)=\sum^{\infty}_{n=0}\frac{1}{n!}\frac{\partial^n}{\partial\lambda^n}K^v_0(\lambda)\bigg | _{\lambda =0 }\lambda^n,
\end{equation}
Substituting equation (\ref{hsnh-02}) into the corresponding symmetry definition equation of the HS-cKdV system (\ref{hs-01})-(\ref{hs-02}), we can conclude that
\begin{equation}\label{hsnh-03}
 K^u_n(\lambda)=\frac{1}{n!}\frac{\partial^n}{\partial\lambda^n}K^u_0(\lambda)\bigg | _{\lambda =0 }\lambda^n,\ \
 K^v_n(\lambda)=\frac{1}{n!}\frac{\partial^n}{\partial\lambda^n}K^v_0(\lambda)\bigg | _{\lambda =0 }\lambda^n,
\end{equation}
must also be a symmetry of the HS-cKdV system for all $n$ $(n=0,1,2...)$. At the same time, we let $\psi_1=\psi_1(\lambda)$ and $\psi_2=\psi_2(\lambda)$
have the formal series form
\begin{equation}\label{hsnh-04}
  \psi_1=\sum^{\infty}_{k=0}\psi_1[k]\lambda^k,\ \ \psi_2=\sum^{\infty}_{k=0}\psi_2[k]\lambda^k,
\end{equation}
where $\psi_1[k]$ and $\psi_2[k]$ are $\lambda$ independent and should be determined later.

Substituting (\ref{hsnh-03}) and (\ref{hsnh-04}) into (\ref{hs-03}) and (\ref{hs-04}), and collecting the coefficients of $\lambda$ yields
\begin{eqnarray}
\label{hsnh-05} &&(\partial^2+u+v)\psi_1[0]=0,\ \ (\partial^2+u+v)\psi_1[k]=-\psi_2[k-1],\\
\label{hsnh-06} &&(\partial^2+u-v)\psi_2[0]=0,\ \ (\partial^2+u-v)\psi_2[k]=\psi_1[k-1].
\end{eqnarray}
Then, $\psi_1[k]$ and $\psi_2[k]$ can be solved recursively as
\begin{eqnarray}\label{hsnh-07}
\hspace{-1.5cm} &&\psi_1[k]=L_1^{-1}\psi_2[k{-}1]=L_1^{-1}L_2^{-1}\psi_1[k{-}2]\cdots=
\begin{cases}(L_2L_1)^{-\frac{k}{2}}\psi_1[0],\ \ k\ \ \mbox{is even} \\
 (L_2L_1)^{{-}\frac{k{-}1}{2}}L^{-1}_1\psi_2[0],\ \ k\ \ \mbox{is odd}
 \end{cases} \\
\hspace{-1.5cm}  &&\psi_2[k]=L_2^{-1}\psi_1[k{-}1]=L_2^{-1}L_1^{-1}\psi_2[k{-}2]\cdots=
\begin{cases} (L_1L_2)^{-\frac{k}{2}}\psi_2[0],\ \ k\ \ \mbox{is even} \\
 (L_1L_2)^{{-}\frac{k{-}1}{2}}L^{-1}_2\psi_1[0],\ \ k\ \ \mbox{is odd}
 \end{cases}\\
 \nonumber \hspace{-1.5cm} &&L_1=-\partial^2-u-v,\ \ L_2=\partial^2+u-v,
\end{eqnarray}
which leads (\ref{hsnh-04}) to
\begin{eqnarray}
\label{hsnh-08}&&\psi_1=\sum^{\infty}_{k_1=0,2,4\cdots}(L_2L_1)^{-\frac{k_1}{2}}\psi_1[0] \lambda^{k_1}+ \sum^{\infty}_{k_2=1,3,5\cdots}(L_2L_1)^{{-}\frac{k_2{-}1}{2}}L^{-1}_1\psi_2[0]\lambda^{k_2},\\
\label{hsnh-09}&&\psi_2=\sum^{\infty}_{k_3=0,2,4\cdots}(L_1L_2)^{-\frac{k_3}{2}}\psi_2[0] \lambda^{k_3}+ \sum^{\infty}_{k_4=1,3,5\cdots}(L_1L_2)^{{-}\frac{k_4{-}1}{2}}L^{-1}_2\psi_1[0]\lambda^{k_4}.
\end{eqnarray}
Finally, the set of the nonlocal symmetries can be
obtained by substituting (\ref{hsnh-08}) and (\ref{hsnh-09}) into (\ref{hsnh-03}):
\begin{eqnarray}
\label{hsnh-10} &&\begin{cases} K^u_n=-2\sum^{n}_{k_1=0,2,4\cdots}(\Xi_1\Xi_{3})_x -2 \sum^{n-1}_{k_2=1,3,5\cdots}(\Xi_2\Xi_{4})_x,\\
 K^v_n=\sum^{n}_{k_1=0,2,4\cdots}(\Xi_1\Xi_{3x}-\Xi_{3}\Xi_{1x}) + \sum^{n-1}_{k_2=1,3,5\cdots}(\Xi_2\Xi_{4x}-\Xi_4\Xi_{2x}),\end{cases} n\ \ \mbox{is even},\\
\label{hsnh-11} && \begin{cases} K^u_n=-2\sum^{n-1}_{k_1=0,2,4\cdots}(\Xi_1\Xi_{6})_x -2 \sum^{n}_{k_2=1,3,5\cdots}(\Xi_2\Xi_{5})_x,\\
 K^v_n=\sum^{n-1}_{k_1=0,2,4\cdots}(\Xi_1\Xi_{6x}-\Xi_{6}\Xi_{1x}) + \sum^{n}_{k_2=1,3,5\cdots}(\Xi_2\Xi_{5x}-\Xi_5\Xi_{2x}),\end{cases} n\ \ \mbox{is odd},
\end{eqnarray}
where
\begin{eqnarray*}
&& \Xi_1=(L_2L_1)^{-\frac{k_1}{2}}\psi_1[0],\ \ \ \ \Xi_2=(L_2L_1)^{{-}\frac{k_2{-}1}{2}}L^{-1}_1\psi_2[0], \\
 && \Xi_3=(L_1L_2)^{-\frac{n-k_1}{2}}\psi_2[0],\ \ \Xi_4=(L_1L_2)^{{-}\frac{n-k_2{-}1}{2}}L^{-1}_2\psi_1[0],\\
&& \Xi_5=(L_1L_2)^{-\frac{n-k_2}{2}}\psi_2[0],\ \ \Xi_6=(L_1L_2)^{{-}\frac{n-k_1{-}1}{2}}L^{-1}_2\psi_1[0].
\end{eqnarray*}

From the set of the nonlocal symmetries, the
negative HS-cKdV hierarchy (the flow equations of the HS-cKdV
equation corresponding to the nonlocal symmetries) follows
immediately:
\begin{eqnarray}\label{hsnh-12}
 &&\begin{cases} u_t= K^u_n,\ \ v_t= K^v_n,\\
 L_1\psi_1[0]=0,\ \ L_2\psi_2[0]=0, \end{cases}
\end{eqnarray}
where $K^u_n$ and $K^v_n$ are expressed by (\ref{hsnh-10}) and (\ref{hsnh-11}).

Using the Miura transformation \cite{j-wj-jmp-1984,j-wj-jmp-1985}
\begin{eqnarray}\label{hsnh-13}
u=-\frac{1}{2}F_{xx}-\frac{1}{4}F^2_x-\frac{1}{4}G^2_x,\ \ v=-\frac{1}{2}G_{xx}-\frac{1}{2}F_x G_x,
\end{eqnarray}
the first one of negative HS-cKdV hierarchy (\ref{hsnh-12}) is transformed to a coupled sinh-Gordon equations:
\begin{eqnarray}\label{hsnh-14}
\nonumber && F_{xt}=\sinh(F)+f,\ \ G_{xt}=\sinh(F)+g,\\
\label{hsnh-14} && [(4s-1)\exp(F)-f]F_x+[\frac{1}{2}\sinh(F)-g]G_x-f_x=0,\\
\nonumber && [f-g-2s\exp(F)](F_x-G_x)+f_x-g_x-2s\exp(F)F_x=0.
\end{eqnarray}
with $s$ is an arbitrary constant. When $s=\frac{1}{4}$ and $G=0$, (\ref{hsnh-14}) is just the usual sinh-Gordon equation.

Further, the complicated integrodifferential
hierarchy (\ref{hsnh-12}) can be written as a simple equivalent
differential equation system $((L_2L_1)^{m+1}P_m=\psi_1[0],(L_1L_2)^mL_1Q_m=\psi_2[0])$
\begin{eqnarray}
 &&\begin{cases} u_t=-2\sum^{m}_{j_1=0}(\Xi_1\Xi_{3})_x -2 \sum^{m-1}_{j_2=0}(\Xi_2\Xi_{4})_x,\\
 v_t=\sum^{m}_{j_1=0}(\Xi_1\Xi_{3x}-\Xi_{3}\Xi_{1x}) + \sum^{m-1}_{j_2=0}(\Xi_2\Xi_{4x}-\Xi_4\Xi_{2x}),\end{cases} n=2m\ \ \mbox{is even},\\
 && \begin{cases} u_t=-2\sum^{m}_{j_1=0}(\Xi_1\Xi_{6})_x -2 \sum^{m}_{j_2=0}(\Xi_2\Xi_{5})_x,\\
 v_t=\sum^{m}_{j_1=0}(\Xi_1\Xi_{6x}-\Xi_{6}\Xi_{1x}) + \sum^{m}_{j_2=0}(\Xi_2\Xi_{5x}-\Xi_5\Xi_{2x}),\end{cases} n=2m+1\ \ \mbox{is odd},\\
 && L_1(L_2L_1)^{m+1}P_m=0,L_2(L_1L_2)^mL_1Q_m=0,
\end{eqnarray}
with
\begin{eqnarray*}
&& \Xi_1=(L_2L_1)^{m-j_1+1}P_m,\ \ \Xi_2=(L_2L_1)^{m-j_2}Q_m, \\
&& \Xi_3=(L_1L_2)^{j_1}L_1Q_m,\ \ \Xi_4=(L_1L_2)^{j_2+1}L_1P_m,\\
&& \Xi_5=(L_1L_2)^{j_2}L_1Q_m,\ \ \Xi_6=(L_1L_2)^{j_1}L_1P_m.
\end{eqnarray*}
In ref.\cite{j-lou-jpa-1997-30}, only the first negative HS-cKdV system is given.
Now we provide the whole negative HS-cKdV hierarchy by simple differential form.

\subsection{Lower-dimensional and Higher-dimensional integrable systems}
From the nonlocal symmetry of the HS-cKdV system and Proposition 5 in Section 2, a nontrivial nonlocal symmetry related to the spectral functions $\psi_1$ and $\psi_2$ can be obtained,
\begin{eqnarray}\label{hse-01}
 \sigma_{N}=(\sigma^u_{N},\sigma^v_{N}) \equiv \Big(-\sum^{N}_{i=1}2a_{i}(\psi_{1i}\psi_{2i})_x, \sum^{N}_{i=1} a_{i}(\psi_{1i}\psi_{2ix}-\psi_{2i}\psi_{1ix})\Big)
\end{eqnarray}
where $a_{i}$ are constants, $i=1,2,...,N$, and $\{\psi_{1i}, \psi_{2i}\}$ are independent solutions of the Lax pair (\ref{hs-03})-(\ref{hs-06}) with $\lambda=0$.

\subsection*{A. Lower-dimensional integrable systems}
Usually, every one symmetry of a higher dimensional model can lead the original one to its
lower form. Now, considering
\begin{equation}\label{hse-02}
  u_x=-\sum^{N}_{i=1}2a_{i}(\psi_{1i}\psi_{2i})_x,\ \ v_x=\sum^{N}_{i=1} a_{i}(\psi_{1i}\psi_{2ix}-\psi_{2i}\psi_{1ix}),
\end{equation}
as a symmetry constraint condition and acting it on the $x$-part of the Lax pair (\ref{hs-03})-(\ref{hs-04}) for $\psi_1=\psi_{1i}$ and $\psi_2=\psi_{2i}$, we
have the lower dimensional 2N-component differential system
%\begin{eqnarray} \label{hse-03}
%\nonumber  && \psi_{1ixx} = 2\sum^{N}_{n=1}a_{n}(\psi_{1n}\psi_{2n})\psi_{1i}-v\psi_{1i}, \\
%           && \psi_{2ixx} = 2\sum^{N}_{n=1}a_{n}(\psi_{1n}\psi_{2n})\psi_{2i}+v\psi_{2i}, \\
%\nonumber  && v_x=\sum^{N}_{n=1} a_{n}(\psi_{1n}\psi_{2nx}-\psi_{2n}\psi_{1nx}).
%\end{eqnarray}
\begin{eqnarray} \label{hse-03}
\nonumber && \psi_{1i}\psi_{1ixxx}-\psi_{1ix}\psi_{1ixx}-\sum^N_{n=1}a_n(\psi_{1n}\psi_{2nx}+3\psi_{2n}\psi_{1nx})\psi^2_{1i}=0,\\
&& \psi_{2i}\psi_{2ixxx}-\psi_{2ix}\psi_{2ixx}-\sum^N_{n=1}a_n(\psi_{2n}\psi_{1nx}+3\psi_{1n}\psi_{2nx})\psi^2_{2i}=0.
\end{eqnarray}
When $N=1$, $\psi_{10}=\psi_1,\psi_{20}=\psi_2$, system (\ref{hse-03}) is the differential equations
\begin{eqnarray}\label{hse-05}
\nonumber && \psi_1\psi_{1xxx}-\psi_{1x}\psi_{1xx}-3a_1\psi^2_1\psi_2\psi_{1x}-a_1\psi^3_1\psi_{2x}=0,\\
&& \psi_2\psi_{2xxx}-\psi_{2x}\psi_{2xx}-3a_1\psi^2_2\psi_1\psi_{2x}-a_1\psi^3_2\psi_{1x}=0,
\end{eqnarray}
More especially, under the condition $\psi_1=\psi_2$, system (\ref{hse-05}) becomes the elliptic equation
\begin{eqnarray*}
\psi^2_{1x}=b_0+b_1\psi^2_1+a_1\psi^4_1,
\end{eqnarray*}
where $b_0$ and $b_1$ are arbitrary constants. So we call system (\ref{hse-05}) a coupled elliptic equation.

Substituting (\ref{hse-02}) into (\ref{hs-05}) and (\ref{hs-06}) and using (\ref{hse-03}), the $t$-part of the Lax pair becomes the generalized 2N-component coupled modified KdV (mKdV) system
%\begin{eqnarray}\label{hse-04}
%\nonumber && \psi_{1it}=-\psi_{1ixxx}+3\sum^{N}_{n=1}a_{n}(\psi_{1n}\psi_{2n})_x\psi_{1i}-3v\psi_{1ix},\\
%&& \psi_{2it}=-\psi_{2ixxx}+3\sum^{N}_{n=1}a_{n}(\psi_{1n}\psi_{2n})_x\psi_{2i}+3v\psi_{2ix},\\
%\nonumber && v_x=\sum^{N}_{n=1} a_{n}(\psi_{1n}\psi_{2nx}-\psi_{2n}\psi_{1nx}).
%\end{eqnarray}
\begin{eqnarray}\label{hse-04}
\nonumber &&\hspace{-1cm} \psi_{1it}=-\psi_{1ixxx}+3\sum^{N}_{n=1}a_{n}(\psi_{1n}\psi_{2n})_x\psi_{1i}-3\sum^{N}_{n=1} a_{n}\partial^{-1}_x(\psi_{1n}\psi_{2nx}-\psi_{2n}\psi_{1nx})\psi_{1ix},\\
&&\hspace{-1cm} \psi_{2it}=-\psi_{2ixxx}+3\sum^{N}_{n=1}a_{n}(\psi_{1n}\psi_{2n})_x\psi_{2i}+3\sum^{N}_{n=1} a_{n}\partial^{-1}_x(\psi_{1n}\psi_{2nx}-\psi_{2n}\psi_{1nx})\psi_{2ix}.
\end{eqnarray}
When $N=1,a_1=1$, $\psi_{10}=\psi_1,\psi_{20}=\psi_2$, system (\ref{hse-04}) is the generalized coupled mKdV equations,
\begin{eqnarray}\label{hse-06}
\nonumber && \psi_{1t}=-\psi_{1xxx}+3\psi_1(\psi_1\psi_2)_x-3v\psi_{1x},\\
&& \psi_{2t}=-\psi_{2xxx}+3\psi_2(\psi_1\psi_2)_x+3v\psi_{2x},\\
\nonumber && v_x=\psi_1\psi_{2x}-\psi_2\psi_{1x}.
\end{eqnarray}

\subsection*{B. Higher-dimensional integrable systems}
To obtain some higher dimensional integrable models, one may introduce some internal parameters.
It is obvious that system (\ref{hs-01})-(\ref{hs-02}) are invariant under the internal parameter translation, say $y$ translation.
That is to say $(u_y, v_y)$ is also a symmetry of the HS-cKdV system.
So we can use
\begin{eqnarray}\label{hse-07}
 u_y=-\sum^{N}_{i=1}2a_{i}(\psi_{1i}\psi_{2i})_x,\ \ v_y=\sum^{N}_{i=1} a_{i}(\psi_{1i}\psi_{2ix}-\psi_{2i}\psi_{1ix}),
\end{eqnarray}
as a generalized symmetry constraint condition.

Substituting (\ref{hse-07}) into the $x$-part of the Lax pair (\ref{hs-03})-(\ref{hs-04}) for $\psi_1=\psi_{1i}$ and $\psi_2=\psi_{2i}$ yields
a higher dimensional 2N-component differential system
\begin{eqnarray}\label{hse-08}
\nonumber  && \psi_{1ixx}-\psi_{1i}\sum^N_{n=1}a_n\partial^{-1}_y(\psi_{1n}\psi_{2nx}+3\psi_{2n}\psi_{1nx})=0, \\
  && \psi_{2ixx}-\psi_{2i}\sum^N_{n=1}a_n\partial^{-1}_y(\psi_{2n}\psi_{1nx}+3\psi_{1n}\psi_{2nx})=0.
\end{eqnarray}
When we take $N=1$, $a_1=1$, $\psi_{10}=\psi_1,\psi_{20}=\psi_2$, and use the same Miura transformation (\ref{hsnh-13}), the system (\ref{hse-08}) also reduces to
the same coupled sinh-Gordon equations with Eqs. (\ref{hsnh-14}).

Considering (\ref{hse-07}) and the $x$-part of the Lax pair (\ref{hs-03})-(\ref{hs-04}) for $\psi_1=\psi_{1i}$ and $\psi_2=\psi_{2i}$ produces
a higher dimensional 2N-component differential system
\begin{eqnarray}\label{hse-09}
\nonumber &&\hspace{-1.5cm}\psi_{1it}=-\psi_{1ixxx}+3\sum^N_{n=1}a_n\partial^{-1}_y(\psi_{1n}\psi_{2n})_{xx}\psi_{1i}-3\sum^N_{n=1}a_n\partial^{-1}_y(\psi_{1n}\psi_{2nx}-\psi_{2n}\psi_{1nx})\psi_{1ix},\\
 &&\hspace{-1.5cm} \psi_{2it}=-\psi_{2ixxx}+3\sum^N_{n=1}a_n\partial^{-1}_y(\psi_{1n}\psi_{2n})_{xx}\psi_{2i}+3\sum^N_{n=1}a_n\partial^{-1}_y(\psi_{1n}\psi_{2nx}-\psi_{2n}\psi_{1nx})\psi_{2ix}.
\end{eqnarray}
Taking $N=1$, $a_1=1$, $\psi_{10}=\psi_1,\psi_{20}=\psi_2$, the system (\ref{hse-09}) becomes
\begin{eqnarray}\label{hse-10}
\nonumber &&\psi_{1t}=-\psi_{1xxx}+3\psi_1\partial^{-1}_y(\psi_1\psi_2)_{xx}-3\psi_{1x}\partial^{-1}_y(\psi_1\psi_{2x}-\psi_2\psi_{1x}),\\
&&\psi_{2t}=-\psi_{2xxx}+3\psi_2\partial^{-1}_y(\psi_1\psi_2)_{xx}+3\psi_{2x}\partial^{-1}_y(\psi_1\psi_{2x}-\psi_2\psi_{1x}).
\end{eqnarray}
For $\psi_1=\psi_2$, $y=x$, system (\ref{hse-10}) reduces to the mKdV equation. So we call system (\ref{hse-10}) a coupled modified ANNV equation.

%coupled ANNV???
%\begin{eqnarray}
%\nonumber &&\psi_{t}=-\psi_{xxx}+3\psi\partial^{-1}_y(\psi^2)_{xx}
%\end{eqnarray}

\section{Conclusions and discussions}
In this paper, we pay close attention to nonlocal symmetries of the HS-cKdV system from DT and their applications.
Some important and meaningful results are obtained.

In general, to search for nonlocal symmetries is an interesting but difficult work. Here, starting from the known DT of the HS-cKdV system,
some different types of nonlocal symmetries are derived directly.
Besides, infinitely many nonlocal symmetries can be obtained by introducing some internal parameters from the seed symmetry.

Our next objective is focused on how to localize the nonlocal symmetry related to the DT.
In fact, for some given nonlocal symmetries of the differential equation(s), whether these symmetries can be transformed to local ones is still unknown.
Fortunately, through introducing five potentials, the nonlocal symmetry obtained from the DT is successfully localized to some local ones in our paper.
This procedure leads the original HS-cKdV system to be extend the prolonged system.
In particular, using the Lie's first theorem to these local symmetries, one can find that the corresponding finite symmetry
transformations have different group parameters from the initial DT but possess the same infinitesimal forms. Meanwhile,
we also observe the fact that the DT of the HS-cKdV system is closely related to the M\"{o}bious transformation of its Schwartz form.

For the prolonged system, the general Lie symmetry transformations and the corresponding similarity reductions are considered.
Some novel exact solutions of the HS-cKdV system are presented, which imply several classes of exact interaction solutions among solitons and other complicated waves including
periodic cnoidal waves, Painlev\'{e} waves and rational waves.
Actually, it is very difficult to obtain these types of solutions from the original DT by solving the spectral problem directly.
The reason lies that to solve the spectral problem with the seed solution being taken non-constant and non-soliton solutions is not an easy work usually.
Therefore, a simple alternative way is provided to construct some new solutions for the integrable models with the known DT.

The left work of this paper
is to extend the HS-cKdV system to some new integrable models from the nonlocal symmetry related to the DT in two aspects.
In fact, the existence of infinitely many symmetries suggests the existence of integrable hierarchies.
By introducing the internal parameter, the negative HS-cKdV hierarchy is obtained without the inverse of the known recursion operator. Using a Miura transformation, this hierarchy
is transformed to a coupled sinh-Gordon hierarchy.
In addition, symmetry constraint approach is one of the most powerful tools to give one new integrable models from known ones. Usually, using this method, one obtains the lower dimensional integrable models from higher ones.
Here, both lower and higher dimensional integrable models are presented by means of the symmetry constraints.

Using the DT to search for nonlocal symmetries of
integrable models and then applying them to construct exact
solutions and new integrable models are both of considerable interest.
However, the concrete integrability for the given lower and higher dimensional models is unknown.
Moreover, in Ref\cite{j-lou-jmp-1997-38}, Lou et al. had made use of the multi-DT to obtain some interesting nonlocal symmetries and constructed various lower and higher dimensional integrable models.
Now a natural problem is how to use those nonlocal symmetries related to the multi-DT to obtain more novel solutions.
These matters are worthy of further study.

\section*{Acknowledgments}
We would like to express our sincere thanks to Professors S Y Lou
and other members of our discussion group for their valuable
comments.
This work is supported by the National Natural Science Foundation of China (Nos. 11275072 and 11075055), Research Fund for the Doctoral Program of Higher Education of China (No. 20120076110024), Innovative Research Team Program of the National Natural Science Foundation of China (No. 61021004), Shanghai Leading Academic Discipline Project (No. B412) and National High Technology Research and Development Program (No. 2011AA010101).

\section*{Acknowledgment}

%% Authors are advised to submit their bibtex database files. They are
%% requested to list a bibtex style file in the manuscript if they do
%% not want to use model1-num-names.bst.

%% References without bibTeX database:

% \begin{thebibliography}{00}

%% \bibitem must have the following form:
%%   \bibitem{key}...
%%

% \bibitem{}

% \end{thebibliography}

\end{document}